\documentclass[aps,preprintnumbers,floats, prd, twocolumn, nofootinbib]{revtex4-1}
\usepackage{graphicx}
\usepackage{dcolumn}
\usepackage{bm}
\usepackage{amsmath}
\usepackage{amsthm}
\usepackage{multirow}
\usepackage{enumerate}
\usepackage{amsfonts}
\usepackage{ifthen}
\usepackage{psfrag}
\usepackage{slashed}
\usepackage{hyperref}
\usepackage{gensymb}
\usepackage[utf8]{inputenc}
\usepackage{color}
\usepackage{subfigure}
\usepackage{ulem}
\usepackage[utf8]{inputenc}

\newcommand{\GeV}{\text{GeV}}
\newcommand{\TeV}{\text{TeV}}

\newcommand{\be}{\begin{equation}}
\newcommand{\ee}{\end{equation}}
\newcommand{\ber}{\begin{eqnarray}}
\newcommand{\eer}{\end{eqnarray}}

\DeclareRobustCommand{\Fig}[1]{Fig.~\ref{#1}}

\DeclareRobustCommand{\Eq}[1]{Eq.~(\ref{#1})}

\DeclareRobustCommand{\Ref}[1]{Ref.~\cite{#1}}
\DeclareRobustCommand{\Refs}[1]{Refs.~\cite{#1}}
\DeclareRobustCommand{\r}[1]{{\rm #1}}

\begin{document} 

\title{Nucleon - Light Dark Matter Annihilation through Baryon Number Violation}

\author{Mingjie Jin$^{1,2}$}
\email{jinmj@ihep.ac.cn}
\author{Yu Gao$^{1}$}
\email{gaoyu@ihep.ac.cn}

\affiliation{$^1$ Key Laboratory of Particle Astrophysics, Institute of High Energy Physics, Chinese Academy of Sciences, Beijing, 100049, China}
\affiliation{$^2$ School of Physical Sciences, University of Chinese Academy of Sciences, Beijing, 100049, China}

\begin{abstract}
Dark matter that participates in baryon-number violating interactions can annihilate with baryons if the dark matter particle is not protected under discrete symmetries. In this paper we investigate the dark matter - baryon annihilation in color-triplet extensions of the Standard Model, in which a fermionic dark matter can become kinematically stable within a small mass range near the proton mass. We demonstrate that the DM's annihilation with nucleons can be probed to stringent limits at large-volume water Cherenkov detectors like the Super-Kamiokande experiment, with the mediator scale $m_\Phi$ constrained up to $10^{7}$\,\GeV. In case of a Majorana light dark matter, this constraint is weaker than, yet close in magnitude to that from neutron-antineutron oscillation. In the Dirac DM case, the dark matter- nucleon annihilation gives stronger bounds than that from the uncertainties of the neutron decay lifetime. 
In a limited range of the DM mass above $m_p+m_e$, the DM-nucleon annihilation bound can be higher than the requirement from the DM's stability in the Universe.
Given the strong limits from Super-Kamiokande, we find it below the current experimental capabilities to detect indirectly the dark matter- nucleon annihilation signal in diffuse Galactic gamma rays or from neutron star heating.
\end{abstract}

\maketitle


\section{Introduction}
\label{sect:intro}

The existence of dark matter (DM) is widely supported by astrophysical~\cite{Zwicky:1937zza, Clowe:2003tk} and cosmological~\cite{Spergel:2006hy, Ade:2015xua} observations. Among many dark matter candidates, a nonthermally produced~\cite{Davoudiasl:2010am, Davoudiasl:2011fj, Allahverdi:2010im, Allahverdi:2013mza, Allahverdi:2017edd} dark matter can participate in baryon number violation ($\Delta B\neq 0$) processes~\cite{Kitano:2004sv, Agashe:2004ci, Cosme:2005sb}. Baryon number violation allows for nucleon destruction and transfers matter-antimatter asymmetry between visible and dark sectors~\cite{Davoudiasl:2010am,Haba:2010bm}, that may generate the correct baryon asymmetry and dark matter abundance at the same time. An interesting aspect of such a light dark matter is that it can stay kinematically stable~\cite{Allahverdi:2013mza,Allahverdi:2017edd, Davoudiasl:2010am} within a narrow mass range close to the proton's mass, where the protection of a discrete symmetry may not be needed\footnote{In the absence of discrete parities, dangerous $LH\chi$ terms can be avoided by introducing new symmetries on leptons.}: a small dark mater - proton mass difference less than the electron mass can avoid the weak decay of dark matter, and also prevent the proton decay via the dark matter's mixing with the neutron through the baryon number violating interaction. Recently this light dark matter has also been studied for its role in a potential semi-invisible decay of the neutron~\cite{McKeen:2015cuz, Fornal:2018eol, Karananas:2018goc}.

In case the dark matter is a Majorana fermion, high-dimension $\Delta B=2$ operators can also be generated, and the current most stringent constraints derive from the Super-Kamiokande (SuperK) measurements on neutron-antineutron oscillation~\cite{Abe:2011ky} and dinucleon decay~\cite{Gustafson:2015qyo}. The resulting bound on the baryon number violation scale is $10^{6-8}$ GeV for operators involving the first two quark generations, and $10\, \TeV$~\cite{McKeen:2015cuz, Allahverdi:2017edd, Dev:2015uca} for the third quark generation. Due to the GeV DM mass and relatively high interaction scale, the cosmic diffuse gamma ray search and nuclear recoil experiments are less effective in probing the light dark matter. Resonance searches at colliders can search for TeV mass mediators~\cite{Dutta:2014kia, Sirunyan:2017jix}.

In addition to the constraints on $\Delta B=2$ processes that require the Majorana nature of the DM fermion,  $\Delta B=1$ operators are available at lower operator dimension no matter whether the dark matter fermion is Majorana or Dirac type. This leads to interesting mixture between dark matter and a neutron, as studied in the neutron decay case~\cite{McKeen:2015cuz, Fornal:2018eol, Karananas:2018goc}. 
In this work we pursue a different perspective of this mixing: the presence of anti-DM (or DM itself in case  it is Majorana) in the Galactic halo allows for the annihilation between the anti-DM and baryons inside terrestrial experiments' detector materials. The annihilation signal is a 2 GeV energy deposit in the form of  multiple light mesons, and flavored mesons can also emerge in case the $\Delta B$ interaction involves heavy quark (HQ) flavors. This final state has been readily searched for at the large-volume water Cherenkov detectors like the SuperK experiment for $n-\bar{n}$ oscillation, and it could also be interpreted into a bound on $\bar{\chi}-N$ annihilation.

In this paper, we adopt the minimal extensions of $SU(3)$ triplet scalar ~\cite{Allahverdi:2013mza, Allahverdi:2017edd,McKeen:2015cuz,Karananas:2018goc, Fornal:2018eol,McKeen:2015cuz, Allahverdi:2017edd, Dev:2015uca,Dutta:2014kia, Allahverdi:2015mha} to the Standard Model (SM) that couples to right-handed quarks and dark matter. We briefly discuss the model layouts in Section~\ref{sec:model} and investigate the phenomena of dark matter - baryon mixing and the annihilation signals in Section~\ref{sec:xsec}.  The experimental sensitivity on annihilation with be given in Section~\ref{sec:limit} and compare with existing limits from other baryon number violating processes. We also discuss the galactic gamma ray signal, and neutron heating limits that arise from dark matter - baryon annihilation in Section~\ref{sec:alter}, and then we conclude in~\ref{sec:conclusion}.


\section{Minimal Standard Model Extensions}
\label{sec:model}

The minimal phenomenological SM  extensions~\cite{Allahverdi:2013mza} for successful baryon number violation and dark matter involve the addition
of heavy $SU(3)$ color-triplet scalar(s) $\Phi$ and a singlet fermionic dark matter $\chi$. Here we consider two such implementations. 

In model I, $\Phi$ is a color-triplet, isospin-singlet scalar $(3,1)_{-1/3}$ with hypercharge $Y=-\frac{1}{3}$. The new interaction Lagrangian terms are
\ber
\mathcal{L}_1&=&\lambda_1\Phi^* \chi d_{R} +\lambda'_1\Phi u_{R}d_{R}+\r{c.c.}\nonumber \\ 
                     &&+m^2_\Phi |\Phi|^2 +\frac{1}{2} m_\chi \bar{\chi}^c\chi ,
\label{eq:lm1}
\eer
where $R$ denote right-handed fermion fields and the color indices are omitted. Here we write a Majorana mass for the dark matter field $\chi$ as the minimal case. Alternative $\chi$ can also be a Dirac fermion, by replacing the $\chi$ with $\chi_R$ at the cost of adding an additional $\chi_L$ to the model. The range of $m_\chi$ is $m_p-m_e<m_\chi<m_p+m_e$ \cite{Allahverdi:2013mza}, where the upper bound to prevent dark matter from decaying to proton and the lower bound from proton stability. One loop correction~\cite{Allahverdi:2013mza} to the $\chi$ mass is  $\delta m_\chi \sim (\lambda/4\pi)^2m_\chi{\rm ln}(\Lambda/m_\Phi)$ where $\Lambda$ is a cut off, and it would not exceed $m_e$ for $\lambda \le \mathcal{O}(0.1)$ with TeV $m_\Phi$.

In the non-thermal relic density production, dark matter and baryons originate from the decay of $\Phi$ fields~\cite{Allahverdi:2013mza}, the couplings $\lambda, \lambda'$ can carry complex phases to generate necessary $CP$ violation in the interference of the  $\Phi\rightarrow q q'$ decay with its self-energy diagrams~\cite{Allahverdi:2010im}. For non-zero $CP$ violation more than one $\Phi$ fields are usually required~\cite{Allahverdi:2013mza}; for dark matter - baryon annihilation at GeV energy, $\Phi$ can be integrated out as heavy field(s), and for convenience we consider the effective $\chi$-quark operators for one such heavy $\Phi$. This can be interpreted as the $\Phi$ with the lowest mass dominates the effective operator, and in the presence of multiple $\Phi$ fields, the total contribution can be proportionally summed up\footnote{Interference is possible when multiple $\Phi$s are mass-degenerate.}.

\begin{figure}[h]
  \centering
    \includegraphics[width=0.32\textwidth]{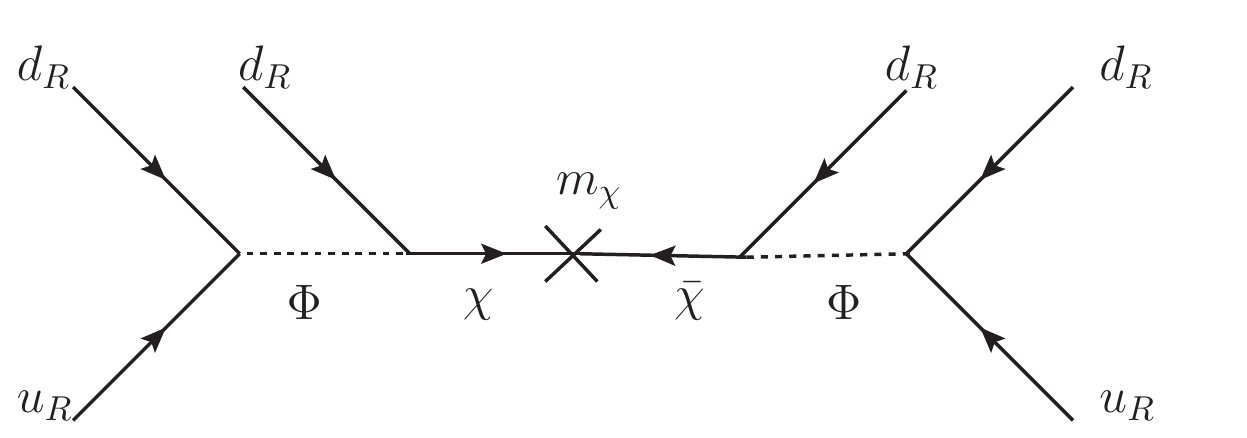}
  \caption{$n-\bar{n}$ oscillation at tree level.}
  \label{fig:nnbartree}
\end{figure}

The dark matter - triquark operator with $\Delta B = 1$ is derived by integrating out the heavy scalar $\Phi$,
\be
\mathcal{L} \supset \frac{\lambda'\lambda}{m^2_{\Phi}} (\chi u_Rd_Rd_R) 
=  \frac{\lambda'\lambda}{m^2_{\Phi}}\cdot \beta (\chi n),
\label{eq:lchiudd}
\ee
where the form factor $\beta$ is defined as $\langle 0|u_{R}d_{R}d_{R} | n\rangle$ and $\beta_{udd} \approx 0.0144~\GeV^3$ from lattice QCD \cite{Aoki:2017puj}. This operator is effectively an off-diagonal mass term between the dark matter and the neutron, leading to a mixing suppressed by $m_\Phi^2$. For a Majorana $\chi$, tree-level $\Delta B = 2$ operators can be generated for $n-\bar{n}$ oscillation, as shown in Fig~\ref{fig:nnbartree}, by repeating the $\Delta B =1$ process with $\chi$'s Majorana mass insertion.  Note in Model I all the three quarks can be valence quarks, thus the light flavor operator is the most stringently constrained by the SuperK $n-\bar{n}$ data. Coupling to heavy quark flavors, in particular with the 3rd generation quarks, can more easily evade the $n-\bar{n}$ oscillation constraints.

In model II, $\Phi=(3,1)_{+2/3}$ couples to two down-type quarks and its hypercharge $Y=+\frac{2}{3}$\,.The  Lagrangian extension is given by
\ber
\mathcal{L}_2&=&\lambda_{2}\Phi^* \chi u_{R}+\lambda'_{2 ij}\Phi d_{Ri}d_{Rj}+\r{c.c.}\nonumber \\
                      & &+m^2_{\Phi} |\Phi|^2+\frac{1}{2} m_\chi \bar{\chi}^c\chi,
\label{eq:lm2}
\eer
where the subscript $i, j$ denote quark flavors. The omitted QCD indexes are antisymmetric with interchanging the two down-type quarks in the second term, thus $d_{Ri},d_{Rj}$ must couple to different quark flavors. Integrating out the $\Phi$ also gives the $\Delta B=1$ operator in \Eq{eq:lchiudd}, and $\chi$ could also be replaced by its right-handed component if $\chi$ is promoted to a Dirac fermion. Unlike in Model I, the $\Delta B =1$ operator will include at least one heavy down-type quark.

\begin{figure}[h]
  \centering
    \includegraphics[width=0.32\textwidth]{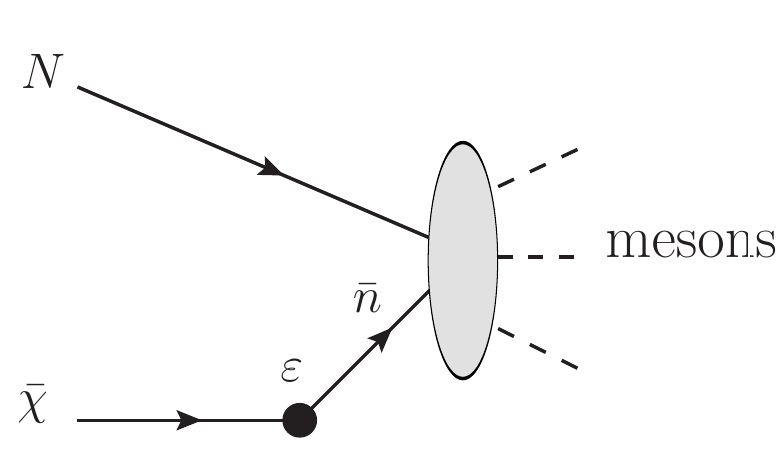}
  \caption{Light anti dark matter annihilates with the nucleon ($N$) through mixing into antineutron $\bar{n}$ and produce a hadronic final state.}
  \label{fig:chinmm}
\end{figure}

The dimension-6 $\Delta B =1$ operator allows the dark matter to mix with neutron ($udd$). Thus the (anti) dark matter ($\chi$) can directly annihilate with the nucleon ($N$) through such mixing and produce a final state of several mesons, as illustrated in \Fig{fig:chinmm}. Due to the close proximity to the proton mass, the final state is almost identical to that of $N\bar{N}$ annihilation.

The dimensional mixing parameter $\varepsilon=\beta\lambda'\lambda/m^2_{\Phi}$ in Fig.~\ref{fig:chinmm} is of mass dimension. The mass eigenstates are then a slight rotation from $\{n,\chi\}$ by $\theta$. In most of the mass range we have $\varepsilon \ll m_n-m_\chi$ (see Appendix \ref{sec:mixangle} for the general case), the mixing angle $\theta$ is
\be
\theta \approx \frac{\varepsilon}{m_n-m_\chi}= \frac{\beta\lambda'\lambda}{m^2_{\Phi}(m_n-m_\chi)},
\label{eq:theta}
\ee
hence the $\chi N$ annihilation cross section can be written in a simple way,
\be
\sigma_{\chi N} \simeq \theta^2 \sigma_{N\bar{n}},
\label{eq:thetasigma}
\ee
up to a tiny shift in the size of phase-space. In the following sections we investigate the signals from this annihilation.


\section{Dark matter - nucleon annihilation}
\label{sec:xsec}

The light dark matter - nucleon annihilation occurs at low relative velocities due to non-relativistic halo dark matter velocity at the Earth. The annihilation cross-section can be inferred from the low-energy cross section of nucleon ($N$) and antinucleon ($\bar{N}$) annihilation~\cite{Dover:1992vj, Zenoni:1999st, Bertin:1997gn, Armstrong:1987nu, Feliciello:1999ti, Mutchler:1988av, Astrua:2002zg, Bertin:1996kw}.  The parametrization of low energy $p,\bar{n}$ annihilation cross section can take the form $\sigma^{\rm ann}_{p\bar{n}}=a+b/P_{\bar{n}}$~\cite{Dover:1980pd}, where $P_{\bar{n}}$ is the antineutron's momentum and the empirical parameters $a,b$ are fitted to experimental data, as given by \Ref{Armstrong:1987nu}, $\sigma^{\rm ann}_{p\bar{n}} =41.4+(29.0\,\r{GeV})/P_{\bar{n}}$ mb in the momentum range 100-500$~\r{MeV}$. 
At low energy, $s$-wave is the most dominant contribution~\cite{Bertin:1997gn, Feliciello:1999ti} for annihilation cross section $\sigma_{p\bar{n}}$, and $\sigma v$ approaches to a constant value $\left.\sigma v\right|_{P_{\bar{n}}=0}=44 \pm 3.5$ mb \cite{Mutchler:1988av} in the low velocity limit.

For the annihilation with one nucleon inside the nucleus ($A$), the $\bar{n}-A$ annihilation cross section has been analyzed with the data for six different nuclei (C, Al, Cu, Ag, Sn and Pb)~\cite{Astrua:2002zg},
\be
\sigma(P_{\bar{n}},A)=\sigma_0(P_{\bar{n}})A^{2/3},
\label{eq:totalxsec}
\ee
where $A$ is the atomic number, and $\sigma_0(P_{\bar{n}})=a+b/P_{\bar{n}}+c/P^2_{\bar{n}}$  (with $a,b,c$ fixed to data). 

As both protons ($p$) and neutrons ($n$) are present inside nuclei, $\sigma_0$ is a weighted average: $\sigma_0=\alpha\sigma_{p\bar{n}}(P_{\bar{n}})+(1-\alpha)\sigma_{n\bar{n}}(P_{\bar{n}})$, where $Z$ is the number of protons and $\alpha=Z/A$. 
Among the tested elements, $\alpha$ varies mildly between 0.4 (lead) and 0.5 (carbon), the total cross-section dependence on $A$ agrees well with ~\Eq{eq:totalxsec} and is insensitive to $\alpha$~\cite{Astrua:2002zg}. Therefore we consider the cross section for both $p-\bar{n}$ and $n-\bar{n}$ to be identical with $\sigma_0$,
and adopt the $\sigma_{p\bar{n}}$ parametrization from \Ref{Armstrong:1987nu}.

As both processes are dominated by strong interaction the $n-\bar{n}$ cross-section is approximately the same as that for $p-\bar{n}$. With $p-\bar{n}$ and $n-\bar{n}$ identical with $\sigma_0$, we then obtain the annihilation cross section for dark matter and a free nucleon (or within a nucleus) at low momentum, by using~\Eq{eq:thetasigma}
\ber
 v\sigma_{\chi N}   &\approx&  44 \times  \frac{(\beta\lambda'\lambda)^2}{m^4_{\Phi}(m_n-m_\chi)^2} \rm{\ mb },\label{eq:sigmavn}\\
  v\sigma_{\chi A}  &\approx&  44 \times  \frac{(\beta\lambda'\lambda)^2 A^{2/3}}{m^4_{\Phi}(m_n-m_\chi)^2} \rm{\ mb }\label{eq:sigmava}.
  \label{eq:sigmav}
\eer


\section{Experimental limits}
\label{sec:limit}

\begin{figure}[h]
  \centering
    \includegraphics[width=0.4\textwidth]{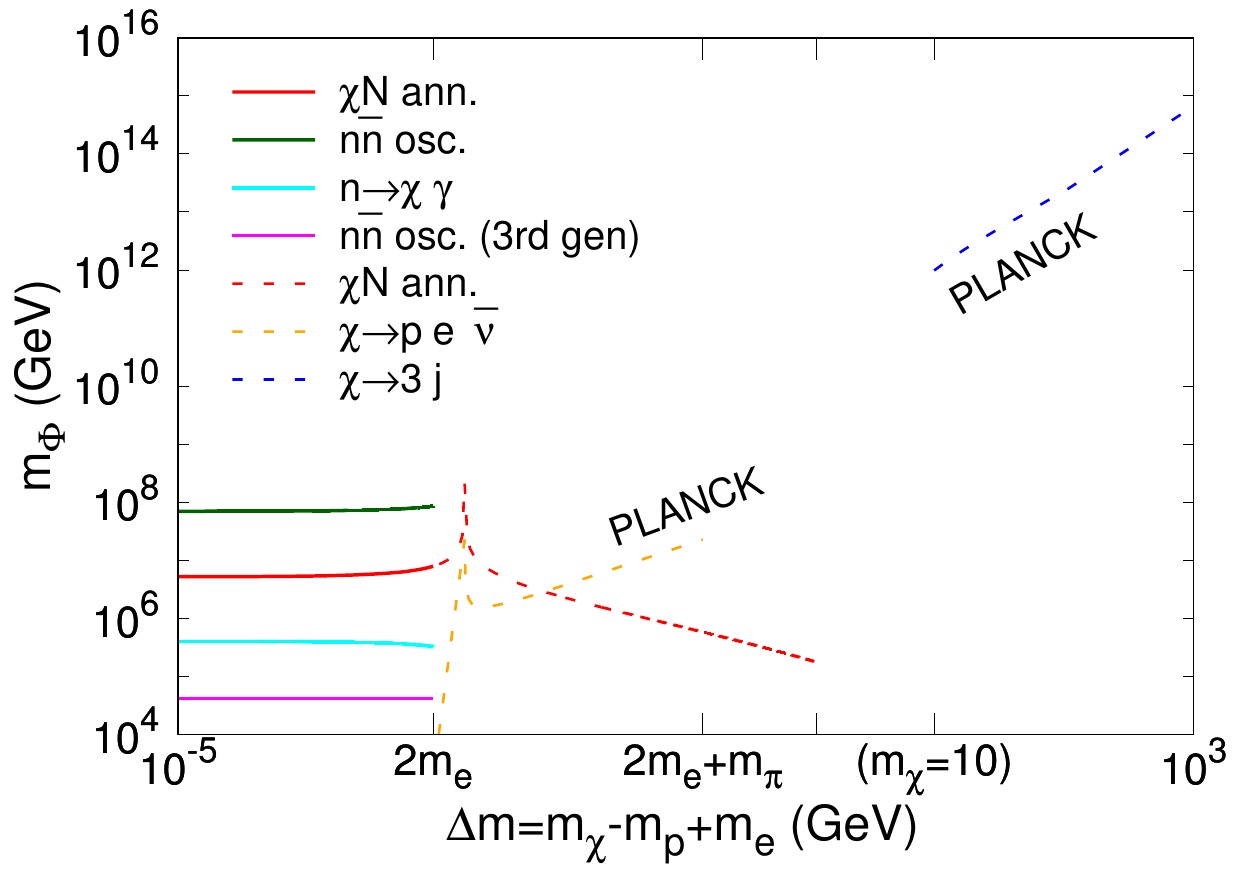} 
  \caption{Experimental limits for $m_\Phi$ versus $\Delta m=m_\chi-m_p+m_e$ from several models. The dark-green, red, cyan, magenta, orange, blue (dashed) lines denote $n-\bar{n}$ oscillation for first generation at tree level, $\chi-N$ annihilation, neutron decays to dark matter, $n-\bar{n}$ oscillation for third generation at loop level, dark matter decays to proton, dark matter decays to three jets where one of three jets is from a $b$-quark,  respectively. The solid  (dashed) line denotes stable (unstable) dark matter. The dark matter can decay when $\Delta m>2m_e$. It only decays to proton when $2m_e<\Delta m<2m_e+m_\pi$ and to three jets when $m_\chi>10\GeV$. The $\chi N$ annihilation bound is extended (dashed) to $m_\chi=2$. All curves assume Majorana $\chi$.}   
 \label{fig:mphi}
\end{figure}

Now we consider the experimental bounds for the $m_\chi$ range above $m_p-m_e$, where the proton decay is kinematically forbidden. In the range $|m_\chi-m_p|<m_e$, $\chi$ is stable, and the $\chi N$ annihilation can be detected at large-volume Cherenkov detectors, along with existing $n-\bar{n}$ oscillation and $n$ decay lifetime limits. For $m_\chi>m_p+m_e$, we will also consider the stability bound of $\chi$ as dark matter, since $\chi$ can decay via mixing to neutrons or directly into 3 quarks/jets at a heavier mass. The $\chi$ decay lifetime will be constrained to $10^{24-26}$s~\cite{Slatyer:2016qyl} by PLANCK~\cite{Ade:2015xua} by the impact on the CMB's propagation. All the experimental bounds are summarized in Fig.~\ref{fig:mphi}. For convenience we give the $m_\Phi$ bounds assuming unity-value couplings $\lambda, \lambda'= 1$. 

The $\chi N$ annihilation leads to a $ \sim\hskip -1mm$ 2\,GeV final state, and it is the same signal as the $n-\bar{n}$ oscillation in which a pair of bound neutrons annihilate inside an oxygen nuclei, in the large volume of water body at SuperK \cite{Abe:2011ky} or SNO \cite{Aharmim:2017jna} experiments. While the $n-\bar{n}$ oscillation rate per unit volume is determined by water density and is subject to a nuclear suppression factor~\cite{Friedman:2008es}, the free-streaming (anti) dark matter directly interacts with nucleons in both hydrogen and oxygen elements and the rate is determined by the local density in the Galactic halo,
%
\be
\frac{{\rm d}N}{{\rm d}t}= A_{\rm eff}\phi_\chi= \eta N_{\rm T} n_\chi \sigma_{\chi {\rm T}}v_\chi
\label{eq:rate}
\ee
where subscript T denotes target nuclei. The `effective area' $A_{\rm eff}=\eta \sigma_{\chi {\rm T}} N_{\rm T}$ is proportional to the detector mass and the experiment's detection efficiency $\eta$. The dark matter flux density is $\phi_\chi=n_\chi v_\chi$ and $\sigma_{\chi {\rm T}}v_\chi$ approaches to the $s-$wave limit at the halo's low $v\sim10^{-3}$ velocity. We take the dark matter number density at the Solar system to be $n_\chi \simeq$ $0.43\,{\rm cm}^{-3} $\,$(\rho=0.4\, \GeV/{\rm cm}^3)$.  For Majorana type of $\chi$, $100\%$ of the local halo density contribution; in the Dirac $\chi$ case, only the anti dark matter $(\bar{\chi})$ component will participate in the $\bar{\chi} N$ annihilation and a reasonable estimate can be placed by assuming $\chi$ and $\bar{\chi}$ each takes 50\% of the halo density. Note that in general the $\chi$ fraction may vary in nonthermal scenarios, from highly suppressed fractions to nearly 100\% in models like the asymmetric dark matter case~\cite{Petraki:2013wwa}.

Here we adopt the SuperK $n-\bar{n}$ oscillation search data \cite{Abe:2011ky} to set the limit on $\chi N$ annihilation. The SuperK contains 22.5 kton fiducial volume of water and the number of target hydrogen (proton) and oxygen from exposure data are $N_p \simeq$ 6.13 $\times 10^{33}$ and $N_{\rm o} \simeq$ 3.06 $\times 10^{33}$\,\cite{Abe:2011ky}. The 24 detected events during a 1489 live-day run is consistent with all-background interpretation~\cite{Abe:2011ky}. 
Here we assume null result and place a 90\% confidence level bound by requiring the $\chi N$ annihilation considering only the statistical uncertainty of the number of background events. With $\eta=12.1\%$, the SuperK constraint on $m_\Phi$ is $\sim \mathcal{O}(10^7)\,\GeV$ in the kinematically stable mass range of Majorana $\chi$, as shown in Fig.~\ref{fig:mphi}, and decreases 16$\%$ for Dirac $\chi$ with a 50\% halo density. The $\chi N$ annihilation bound is also extended into the unstable range $(m_\chi > m_p + m_e)$, where the shape of the constraint curve is determined by the $m_\chi$ dependence of the $\chi n$ mixing angle $\theta$. The sharp rise at $m_\chi = m_n$ is due to the vanishing of the mass difference which makes $\theta$ less sensitive to $m_\Phi$ for $\varepsilon$ comparable or larger than $|m_\chi-m_n|$.

For Majorana $\chi$, $\Delta B=2$ operators lead to $n-\bar{n}$ oscillation, and \Refs{McKeen:2015cuz, Allahverdi:2017edd, Dev:2015uca} studied the corresponding constraints. For $n-\bar{n}$ oscillation with first generation at tree level and third generation at one loop level, the free $n-\bar{n}$ oscillation time is given in Ref.~\cite{Allahverdi:2017edd} and~\cite{McKeen:2015cuz}

\be
 \tau^{-1}_{n\bar{n}} \simeq \left\{
              \begin{array}{ll}
               \frac{\beta^2\lambda'^2_1\lambda^2_1 m_\chi}{m^4_\Phi(m^2_n-m^2_\chi)},  & ({\rm 1st\,gen.}) \\
               & \\
               \frac{\lambda'^4_2\lambda^2_2 m_\chi}{16\pi^2m^6_\Phi} \Lambda^6_{\rm QCD}{\rm ln}\left(\frac{m^2_\Phi}{m^2_\chi}\right),  & ({\rm 3rd\,gen.})
               \end{array}         
               \right.             
\ee
where $\Lambda_{\rm QCD}^6$ is from the MIT bag model \cite{Rao:1983sd, Babu:2001qr}. 
The SuperK lifetime limit is $\tau_{n\bar{n}} > 2.7\times 10^8 \r{s}$~\cite{Abe:2011ky}, where the first generation gives the strongest bound on in \Fig{fig:mphi} than $\chi N$ annihilation, as shown in Fig.~\ref{fig:mphi}.

A Dirac-type $\chi$ does not generate $n-\bar{n}$ oscillation and evades the stringent $n-\bar{n}$ oscillation and dinucleon decay bounds. Yet the $\Delta B =1$ operator still exists, and $\bar{\chi} N$ annihilation, $\chi$ stability constraints apply. Here we compare the $\bar{\chi} N$ annihilation results with those from $\chi$ decay limits. The dark matter decay process $\chi \rightarrow pe\bar{\nu}$ occurs via the $\chi \bar{n}$ mixing, and the four-fermion interaction is simply scaled by the mixing angle $\theta$ from the SM's effective weak interaction operator for $n\rightarrow p e\bar{\nu}$ decay. For the mass range $m_p+m_e<m_\chi<m_p+m_e+m_\pi$ we compute the $\chi$ decay by phase space since there is no other significant final state. More complicated final states will rise for $m_\chi>m_p+m_e+m_\pi$ and a full study for heavier $\chi$ decay with a multiplicity in mesons is beyond the main focus of this paper. Instead, we present the $\chi\rightarrow qqq$ decay limits for a higher mass range $m_\chi > {\cal O}(10)$ GeV, where each quark hadronizes individually into jet(s). The stability bounds are shown by the orange and blue dashed curves in~\Fig{fig:mphi}.

The $\chi$ decay leads to gamma ray signals from final state $\pi^0$ decay that can be probed by Fermi-LAT diffuse gamma ray searches~\cite{Baring:2015sza}, and also by CMB polarization data on the heating effect of the decay-injected photons and charged particles~\cite{Ade:2015xua}. With $m_\chi$ below $m_p+m_e+m_\pi$, PLANCK gives leading constraint $\tau_{\chi}>10^{24}$s\,\cite{Clark:2018ghm}. For heavy $\chi\rightarrow 3j$ decay, the Fermi-LAT limit for the hadronic decay channel is $\tau_{\chi}>10^{26}{\rm s}$\,\cite{Baring:2015sza} that is more stringent than the PLANCK bound $\tau_{\chi}>10^{24}{\rm s}$ \cite{Clark:2018ghm, Slatyer:2016qyl}. Note the in the $\chi \rightarrow pe\bar{\nu}$ channel the electron energy is only a small fraction of the $m_\chi$, the PLANCK bound is scaled proportionally by the energy injection rate to the ${\rm DM}\rightarrow e^+e^-$ limit at $m_{\rm DM}\sim 2E_e$.

Recently Ref.~\cite{Fornal:2018eol} studied the neutron's semi-invisible decay $n\rightarrow \chi \gamma$ in the Dirac $\chi$ case to interpret the neutron decay anomaly~\cite{Serebrov:2017bzo, Pattie:2017vsj, Yue:2013qrc}, which gives $|\lambda\lambda'|/m^2_\Phi$ $\approx 6.7 \times 10^{-6}\, {\rm TeV}^{-2}$ at $m_\chi$=0.9379 MeV. For comparison, Fig.~\ref{fig:mphi} illustrates this scenario as the cyan curve, and we find it constrained by the $\chi N$ annihilation bound.

A few notes are due about the heavy quark flavor cases in \Eq{eq:lchiudd} that are typical in Model II. By replacing one light quark with a flavored one, \Eq{eq:lchiudd} can mediate $\Delta s$, $\Delta c$, $\Delta b\neq 0$ processes if kinematically permitted. For 1 GeV Majorana dark matter,  $D, B$ mesons are too massive to be produced from low energy $\chi N, NN$ interactions. However,  $K$ meson production is kinematically allowed and the dinucleon decay $NN\rightarrow KK$ via dimension-9 $\Delta s=2$ operator is constrained to $m_\Phi> 10^7\GeV$ by SuperK measurement~\cite{Gustafson:2015qyo}. Heavy flavored operators can also lead  to $\chi - n$ mixing via the presence of  heavy quark inside a neutron, namely considering the effective tri-quark form factors $\beta_c, \beta_s, \beta_b$ analogous to the $\beta$ in \Eq{eq:theta}. Since the $\chi N$ annihilation cross-section scales as $\sigma\propto \beta^2m_\Phi^{-4}$, the SuperK limit on $m_\Phi$ scales $\beta_{\rm HQ}^{-1/2}$ for heavy quark flavor cases. Such form factors are currently unavailable and their evaluation in future lattice QCD studies will be greatly helpful.

\begin{figure}[h]
  \centering
   \subfigure[~$n+\chi \to \pi+\pi$]{
      \label{subfig:pipi} 
   \includegraphics[width=0.32\textwidth]{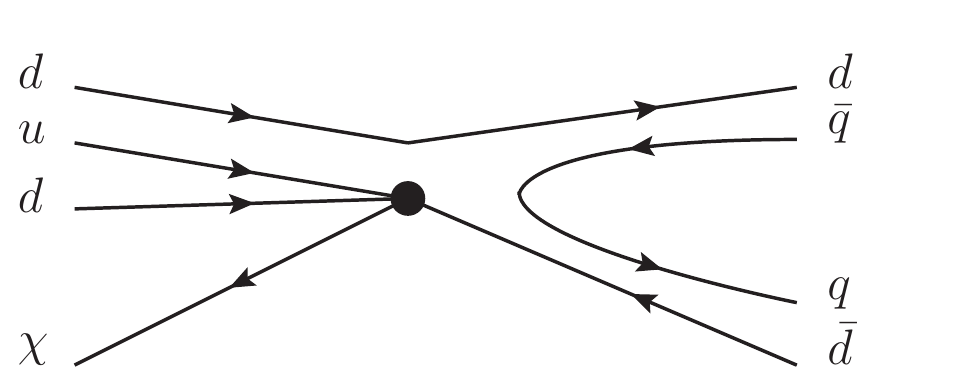}}%
  \hspace{0.01\textwidth} 
  \subfigure[~$n+\chi \to \pi+K$]{ 
     \label{subfig:pik} 
    \includegraphics[width=0.32\textwidth]{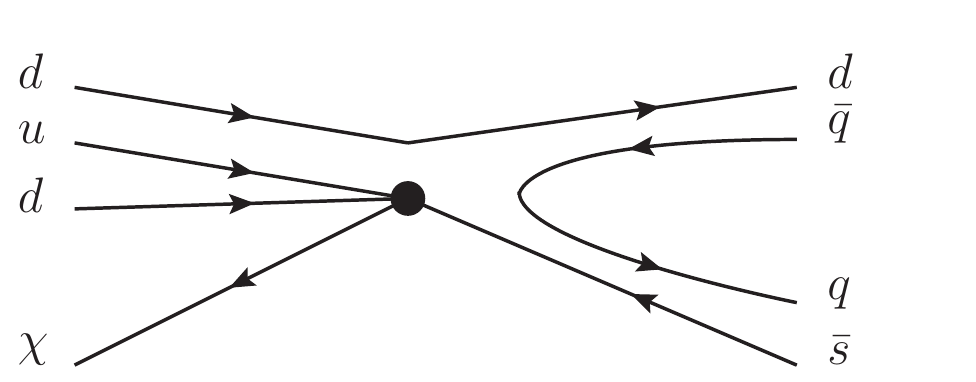}} 
  \caption{The dark matter interacts with neutron and produces two mesons. Two processes with different down-type quark: $n+\chi \to \pi+\pi$ and $n+\chi \to \pi+K$.}
  \label{fig:pchi}
\end{figure}

Still, one can circumvent the $\beta$ form-factor suppression of the neutron's heavy quark presence in a few ways. Ref.~\cite{Allahverdi:2017edd} proposed a loop-level $\Delta b =2$ process where the heavy (b) quark hides as internal lines and the external fermions only involve valence quarks. The corresponding $n-\bar{n}$ oscillation bound is $m_\Phi>10^4\GeV$, as shown in Fig.~\ref{fig:mphi} (magenta). Note by inciting the $\beta$ form factor we assume the heavy fermion presence in the initial state of $N\bar{N}\rightarrow \pi(s)$ annihilation. However, this is unnecessary as the heavy quark can also be the out-going fermion, as illustrated in Fig. ~\ref{fig:pchi}, where the $\chi N \rightarrow \pi K$ does not suffer from an initial-state neutron's $\beta_s$. The two annihilation processes in \Eq{fig:pchi} are kinematically similar, and we expect $\sigma_{\pi K}$ from the $\Delta s=1$ operator to be comparable to a $\sigma_{\pi \pi}$ generated by valence-quark-only operators, up to a reduction in the final state phase space due to the $K,\pi$ mass difference. While the SuperK dinucleon decay data focused on pair-produced $K$ meson signals,  we can still estimate the bound on  $NN\rightarrow\pi K$ by scaling with the kaon multiplicity and obtain an $m_\Phi\sim \mathcal{O}(10^{6-7})\,\GeV$ bound for the $\Delta s=1$ operator. Additionally, one may consider operators by replacing more than one valence quark with heavy quarks in \Eq{eq:lchiudd}, further reducing their constraints from $\chi N, n-\bar{n}, NN$ interactions. For baryon number generation at high energy scale, heavy-flavored operators work equally well as 1st generation quark operators, plus characteristic flavored signals in high-energy collider searches.

\section{Alternative searches}
\label{sec:alter}

Hydrogen and helium are the two most abundant components that make up nearly 99$\%$ \cite{Ferriere:2001rg} of the interstellar gas in our galaxy. $\chi N$ annihilation occurs between dark matter and the galactic gas and produces a hadronic final state, where the $\pi^0$ decay may yield a diffuse photon signal. 
The photon signal intensity is written in a similar way as that from dark matter annihilations,
\ber
&& \frac{{\rm d} \phi_\gamma(\chi N)}{ {\rm d} E {\rm d} \Omega} = \theta^2 \frac{\langle v\sigma_{\chi N} \rangle}{8\pi m_\chi m_N}  \frac{{\rm d} N_\gamma}{{\rm d} E} \int_{\rm los} \rho_\chi \rho_B {\rm d}s\,,\\
&&\frac{{\rm d} \phi_\gamma(\chi\chi)}{ {\rm d} E {\rm d} \Omega} = \frac{\langle v\sigma_{\chi\chi} \rangle}{8\pi m^2_\chi}  \frac{{\rm d} N_\gamma}{{\rm d} E} \int_{\rm los}  \rho^2_\chi {\rm d}s\,, \label{eq:chichibar}
\eer
where the integration over density profiles is along the line-of-sight (los) direction. $\langle \sigma v \rangle$ is the thermally averaged annihilation cross section, and ${\rm d} N_\gamma/ {\rm d} E$ is the final state photon spectrum. Here we take $\langle \sigma v \rangle_{\chi N} \approx \sigma v_{\chi N}$ for $s-$wave dominated annihilation at low velocity $v\sim 10^{-3}$. The DM density $\rho_\chi$ follows the halo distribution, e.g. the NFW profile $\rho_\chi=\rho_s (r/r_s)^{-1} (1+r/r_s)^{-2}$~\cite{Navarro:1996gj}, where $\rho_s=0.33\,\GeV/ {\rm cm}^3$ and  $r_s=20$ kpc. For the Galactic diffuse gas distribution, we assume a simple hydrogen density parametrization $n_{\rm gas}=0.0135(r/1{\rm kpc})^{-1.5} {\rm cm}^{-3}$ in Ref.~\cite{0004-637X-800-1-14}. Note the diffuse gas density is typically two orders of magnitude below the DM density. Given the SuperK limit $m_\Phi$ $\ge$ $\mathcal{O}(10^{7})$\,\GeV, the annihilation cross-section $\langle \sigma v \rangle_{\chi N} \le \mathcal{O}(10^{-41})$\,cm$^{3}\,{\rm s}^{-1}$. Thus the $\chi N$ signal flux is significantly below the Fermi-LAT's sensitivity~\cite{Ackermann:2015lka} of $10^{-30} \,{\rm cm}^{3}{\rm s^{-1}}$ at 1 GeV DM-DM annihilation.

\medskip

Dark matter - nucleon annihilation can lead to direct energy deposit into stars. In addition, the $\Phi^* u_R\chi$ term leads to $\chi-N$ scattering that allows for stars to capture dark matter~\cite{Gould:1987ir}, and it is not suppressed by $\chi-N$ mixing angle.  Both the free-streaming dark matter that transverse stars, as well as those gravitationally captured, may annihilate with stellar nucleons and provide a steady heat source to cold and dense systems like aged neutron stars~\cite{Ozel:2016oaf}. A typical neutron star of $\sim 1.5$ solar mass and 10 Km radius is opaque to the dark matter flow for an $\chi N$ annihilation/scattering cross-section greater than a geometric threshold $\sigma_{\rm th}=10^{-45}$cm$^2$ for $m_{\rm DM}$ between 1 GeV and $10^6$ GeV~\cite{Baryakhtar:2017dbj}. Note that in our case $\chi$ annihilates with a nucleon and there is greater amount of energy release per unit DM mass if compared to conventional DM-DM annihilation cases. The energy deposit thermalizes with the neutron star and is eventually radiated away, raising the neutron star's surface temperature. For recent DM-induced neutron star heating studies, see~\Refs{Kouvaris:2010vv, deLavallaz:2010wp, Raj:2017wrv, Bell:2018pkk}. The SuperK's $\chi N$ bound requires $\theta^2 \leq 10^{-26}$ and the direct $\chi N$ annihilation is less than $10^{-51}$cm$^2$. The average DM flux density through the neutron star is
\be 
\frac{{\rm d}N_{\rm ann}}{{\rm d}t}=  N_{\rm NS} n_\chi \sigma_{\chi N}v_\chi
\ee
where $N_{\rm NS}=M_{\rm NS}/m_n$ is the total number of neutrons in the neutron star. Due to gravitational acceleration, the DM gains an average of $35\%$ kinetic energy after falling into the neutron star, and with $v\sim0.67$ the $\chi N$ annihilation cross-section can still be well describe by $\sigma^{\rm ann}_{\chi n} = \theta^2\sigma^{\rm ann}_{n\bar{n}}$, where the parametrization is given by $\sigma^{\rm ann}_{n\bar{n}}=38.0+35.0/P_{\bar{n}}(\GeV)$~\cite{Mutchler:1988av}. Then the direct annihilation rate is $\sim 10^{16}\,{\rm s^{-1}}$ for uncaptured DM that transverse the neutron star. Since the annihilation rate is far below saturation, DM capture by elastic scattering would also contribute to the heating process.
 
\Ref{Allahverdi:2013mza} gave the TeV scale $\Phi$ mediated  $\chi N$ elastic  cross section as  $\sigma_{\rm SI} \le 10^{-(15-16)}$ pb for spin-independent (SI) scattering and $\sigma_{\rm SD} \le 10^{-(5-6)}$ pb for spin-dependent (SD) scattering.  $\sigma_{\rm SI}$ is suppressed by $m^8_\Phi$ and becomes subdominant at large $m_\Phi$. In contrast $\sigma_{\rm SD}$ from the leading $(\bar{\chi}\gamma_5\gamma^\mu\chi)(\bar{q}\gamma_5\gamma_\mu q)/m^2_\Phi$ operator scales with $m^{-4}_\Phi$. For $\sigma_{\rm ela} \ll \sigma_{\rm th}$ the capture rate by neutron star is proportional to $\sigma_{\rm ela}/\sigma_{\rm th}$ and with $m_\Phi=10^{7}$ GeV the SD scattering leads to $\sim 10^{10}\,{\rm s^{-1}}$. This is still lower in comparison with direct annihilation, and the energy deposit is dominated by direct $\chi n$ annihilation.  

The heating rate is $\dot{E}={E_{\rm t} \dot{N}}f$~\cite{Baryakhtar:2017dbj}, where $\dot{N}=\pi b^2v_\chi n_\chi$ is the number rate of dark matter flux and $b$ is the impact parameter, $f=\sigma_{\chi N}^{\rm ann}/\sigma_{\rm th}$ is the annihilation efficiency, and each $\chi$ contributes energy equal to $E_{\rm t}=2.35 m_\chi$. A neutron-star sized black body's surface temperature is $T_{\rm s}\simeq 134\,{\rm K}\,(10^7\,\GeV/m_\Phi)$ at $m_\chi=m_p+m_e$ by assuming equilibrium between the DM heating and black-body radiation. To a distant observer this temperature lowers to $T\simeq 100\,{\rm K}\,(10^7\,\GeV/m_\Phi)$ due to gravitational redshift, and is below the current experimental sensitivities~\cite{jwst}.

\section{Conclusion}
\label{sec:conclusion}

In this paper we investigated the dark matter annihilate with baryons (nucleons) through baryon number violating extensions to the Standard Model. For these models we assume color-triplet, iso-singlet scalar(s) and a fermionic dark matter, in which the dark matter is kinematically stable within a small mass range near the proton mass.

The annihilation between dark matter and nucleon depends the on the DM-neutron mixing angle, and the cross-section can be determined from the nucleon annihilation cross-section measurements. The mixing angle given in Eq.~\ref{eq:theta} applies to both Dirac and Majorana fermion DM cases, and it provides good probe to the scale of the mediator $m_\Phi$. Due to the identical final state between the $n-\bar{n}$ oscillation and $\chi N$ annihilation at the SuperK experiment, we constrain the stringent limit $m_\Phi$ up to $10^{7}$\,\GeV \hskip 0.9mm by re-interpreting the $n-\bar{n}$ oscillation data for the DM - nucleon annihilation. For Majorana-type dark matter, the constraint is one order in magnitude lower than the bound from neutron-antineutron oscillation. In the Dirac case,  DM - nucleon annihilation gives much stronger bounds than that from neutron decay lifetime uncertainties. We also extended the SuperK bound into the mass range $m_\chi> m_p+m_e$ and compare with the DM stability limits. For a small mass range as illustrated in \Fig{fig:mphi}, the SuperK bounds exceed that from DM stability.

For operators involving heavy quarks, DM - nucleon annihilation occurs either via the heavy quark's presence in neutrons, or through the unsuppressed $\Delta s = 1$ annihilation processes as the $K$ meson is kinematically allowed in the 2 GeV final state. The $\Delta s =1$ annihilation can also be strongly constrained by SuperK dinucleon data.

We then discussed the prospects of dark matter- nucleon annihilation in the indirect detection of the Galactic diffuse gamma rays, and in neutron star heating. Given that the $m_\Phi$ is severely constrained by SuperK, 
a $\langle \sigma v \rangle_{\chi N} \le \mathcal{O}(10^{-41})$\,cm$^{3}\,{\rm s}^{-1}$ for indirect detection, or a heated neutron star temperature  $T\simeq 100\,{\rm K}$ are significantly below the reach of current experiments.

\bigskip
{\bf Acknowledgements}

\medskip
The authors thank Xiao-jun Bi, Hong-bo Hu and Eigo Shintani for helpful discussions. Y.G. is supported under grant no.~Y7515560U1 by the Institute of High Energy Physics, Chinese Academy of Sciences. M.J. thanks the Institute of High Energy Physics, Chinese Academy of Sciences for support.


\appendix

\section{Dark matter-neutron mixing angle}
\label{sec:mixangle}

\begin{figure}[h]
  \centering
    \includegraphics[width=0.32\textwidth]{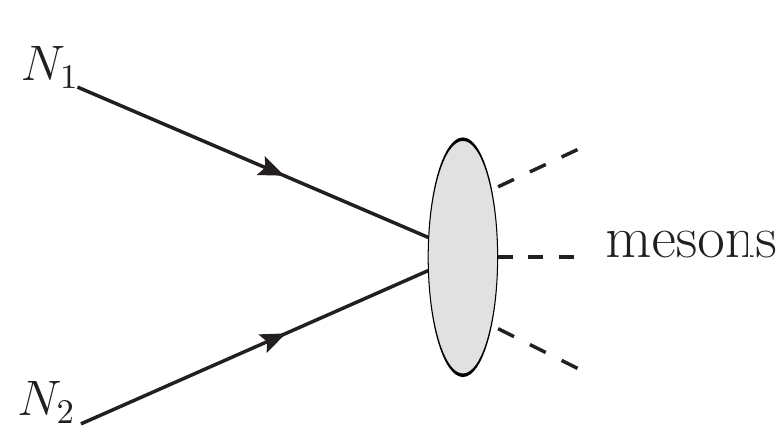} 
  \caption{Two mass eigenstates $N_1$ and $N_2$ annihilate to several mesons.}
  \label{fig:n1n2mm}
\end{figure}

The Lagrangian for dark matter and neutron ($\chi-n$) interaction can be written by
\be
\mathcal{L}_{\chi n}=m_nn\bar{n}+m_\chi \chi\chi+\varepsilon n\chi+\cdots,
\label{eq:lchin}
\ee
where $m_\chi$ and $m_n$ are dark matter and neutron masses, the mixing parameter in Eq.\,(\ref{eq:lchiudd})
$\varepsilon=\beta\lambda'\lambda/m^2_{\Phi}$ and the ellipsis denote kinematic terms. We set $N_1, N_2$ are mass eigenstates, as is illustrated in Fig. \ref{fig:n1n2mm}, and

\ber
    \left(
 \begin{matrix}
   N_1 \\
   N_2
  \end{matrix}
  \right)
 & =& U
    \left(
 \begin{matrix}
   \chi \\
   n
  \end{matrix}
  \right)\nonumber\\
  &=&
\left(
 \begin{matrix}
{\rm cos}\,\theta  &{\rm sin}\,\theta \\
-{\rm sin}\,\theta  & {\rm cos}\,\theta
  \end{matrix}
  \right)  
  \left(
 \begin{matrix}
   \chi \\
   n
  \end{matrix}
  \right).
  \label{eq:matrixchin}
  \eer

The mixing matrix $U$ can be written by using the \Eq{eq:lchin} 

\be
U= C
\left(
 \begin{matrix}
1  & \frac{-\Delta+\sqrt{\Delta^2+4\varepsilon^2}}{2\varepsilon} \\
- \frac{-\Delta+\sqrt{\Delta^2+4\varepsilon^2}}{2\varepsilon} & 1
  \end{matrix}
  \right),  
  \ee
where $\Delta=m_n-m_\chi>0$ and $C$ is a normalization constant. The mixing angle can be written as 
\be
\theta={\rm arctan}\left(\frac{-\Delta+\sqrt{\Delta^2+4\varepsilon^2}}{2\varepsilon}\right). 
\ee

In most of the parameter space, $\varepsilon /(m_n-m_\chi) \ll 1$, the mixing angle is 
\be
\theta \simeq \frac{\varepsilon}{m_{\bar{n}}-m_\chi},
\ee
and then
\be
U \simeq
\left(
 \begin{matrix}
1  & \frac{\varepsilon}{m_{\bar{n}}-m_\chi} \\
- \frac{\varepsilon}{m_{\bar{n}}-m_\chi} & 1
  \end{matrix}
  \right).    
\ee
For $\varepsilon /(m_n-m_\chi) \gg 1$, the $\theta$ approaches to $45^\circ$ and

\be
U \simeq
\frac{1}{\sqrt{2}}\left(
 \begin{matrix}
1  & 1 \\
- 1 & 1
  \end{matrix}
  \right),    
\ee
which becomes maximal mixing and insensitive to $\varepsilon$.

\bibliography{bibfile}

\end{document}